# PYROELECTRIC HARVESTING FROM STEADY HEAT SOURCES VIA ACTIVE PIEZOELECTRIC THERMAL SWITCHING

*Jun Usami*, Yuki Okamoto*, Shimpei Saito, Saeko Tachikawa, Soumei Baba, Takeshi Fujii, Sucheta Gorwadkar, Natsumi Makimoto, and Takeshi Kobayashi*
National Institute of Advanced Industrial Science and Technology (AIST), JAPAN

## ABSTRACT

This study demonstrates pyroelectric energy harvesting from a steady heat source using active piezoelectric thermal switching in a PZT/Si bimorph cantilever. The same PZT layer serves as both the mechanical actuator and the pyroelectric energy-conversion element. Dual output electrodes enable separation of the pyroelectric and piezoelectric signals. Pyroelectric energy harvesting is experimentally demonstrated at frequencies up to 110 Hz and temperature differences as small as 5 K. Comparison with the response predicted from the independently measured pyroelectric coefficient, together with transient thermal simulations, identifies dynamic interfacial heat transfer as a key limitation to high-frequency operation. This work demonstrates a compact MEMS approach for higher-frequency pyroelectric energy harvesting from low-grade steady heat sources.

## KEYWORDS

Thermal energy harvesting, Pyroelectric effect, Piezoelectric thermal switching, Waste heat recovery

## INTRODUCTION

Low-grade heat is widely available in the environment and from various devices, yet its effective utilization remains challenging because of its relatively low temperature and small temperature difference from the surroundings. Thermoelectric generators, which have been widely investigated as solid-state energy-conversion devices, also suffer from a rapid reduction in output power at small temperature differences, as their maximum output power under matched-load conditions scales approximately with the square of the temperature difference [1]. In contrast, pyroelectric energy conversion generates electrical energy from temporal temperature variations of the material, providing an alternative approach for harvesting energy from low-temperature, small-temperature-difference heat sources [2].

Pyroelectric energy conversion inherently requires temporal temperature variation. One practical application is the direct harvesting of naturally fluctuating heat sources, such as human respiration [3]. Continuous energy harvesting from more general steady heat sources, in contrast, requires integration with a mechanism that periodically changes the temperature of the pyroelectric material. For this purpose, passive approaches that generate self-sustained temperature cycles using thermomechanical structures have been proposed and experimentally demonstrated [4]–[6]. These passive systems have typically required temperature differences of several tens of Kelvin or more. Externally driven thermal-cycling approaches have also demonstrated high pyroelectric energy-harvesting performance, highlighting their potential for efficient heat-to-electricity conversion [2], [7], [8]. Both approaches require additional mechanisms to generate the temperature cycles and, more importantly, have generally been limited to relatively slow thermal-cycling frequencies of approximately 1 Hz or lower, as summarized in Fig. 1. The relatively low operating frequencies can be attributed, at least in part, to the thermal relaxation time of the pyroelectric materials and device structures, as well as the response time of the thermal-switching mechanisms. For a given energy conversion per cycle, the average harvested power increases with operating frequency, making higher-frequency operation important for increasing pyroelectric power output. However, pyroelectric energy harvesting based on high-frequency temperature modulation generated from a steady heat source has not yet been sufficiently demonstrated experimentally, and the factors limiting such high-frequency operation remain unclear.

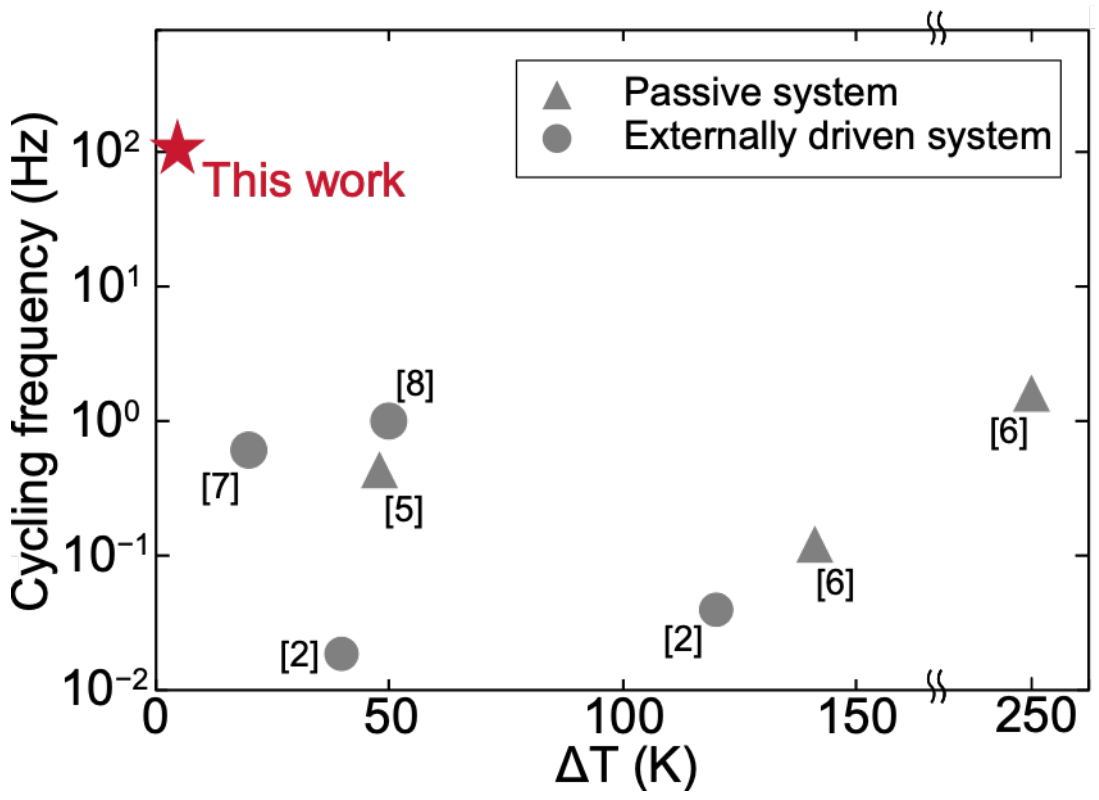

*Figure 1: Comparison of thermal cycling frequency and temperature difference between the heat source and sink for previously reported pyroelectric energy-harvesting systems [2], [5]–[8] and this work.*

In this study, we demonstrate pyroelectric energy harvesting from a steady heat source using an active MEMS thermal switch that generates periodic temperature modulation. To extend the operation toward higher frequencies, a compact thin-film MEMS structure was employed to achieve both a short thermal relaxation time and a fast mechanical response. In addition, a single lead zirconate titanate (PZT) film serves as both the piezoelectric actuator and the pyroelectric energy-conversion element, enabling thermal switching and energy conversion in a simple integrated structure without requiring a separate actuator. Experimentally, the pyroelectric response was separated from the simultaneously generated piezoelectric signal, allowing pyroelectric operation to be verified at temperature differences down to 5 K and at frequencies up to 110 Hz.

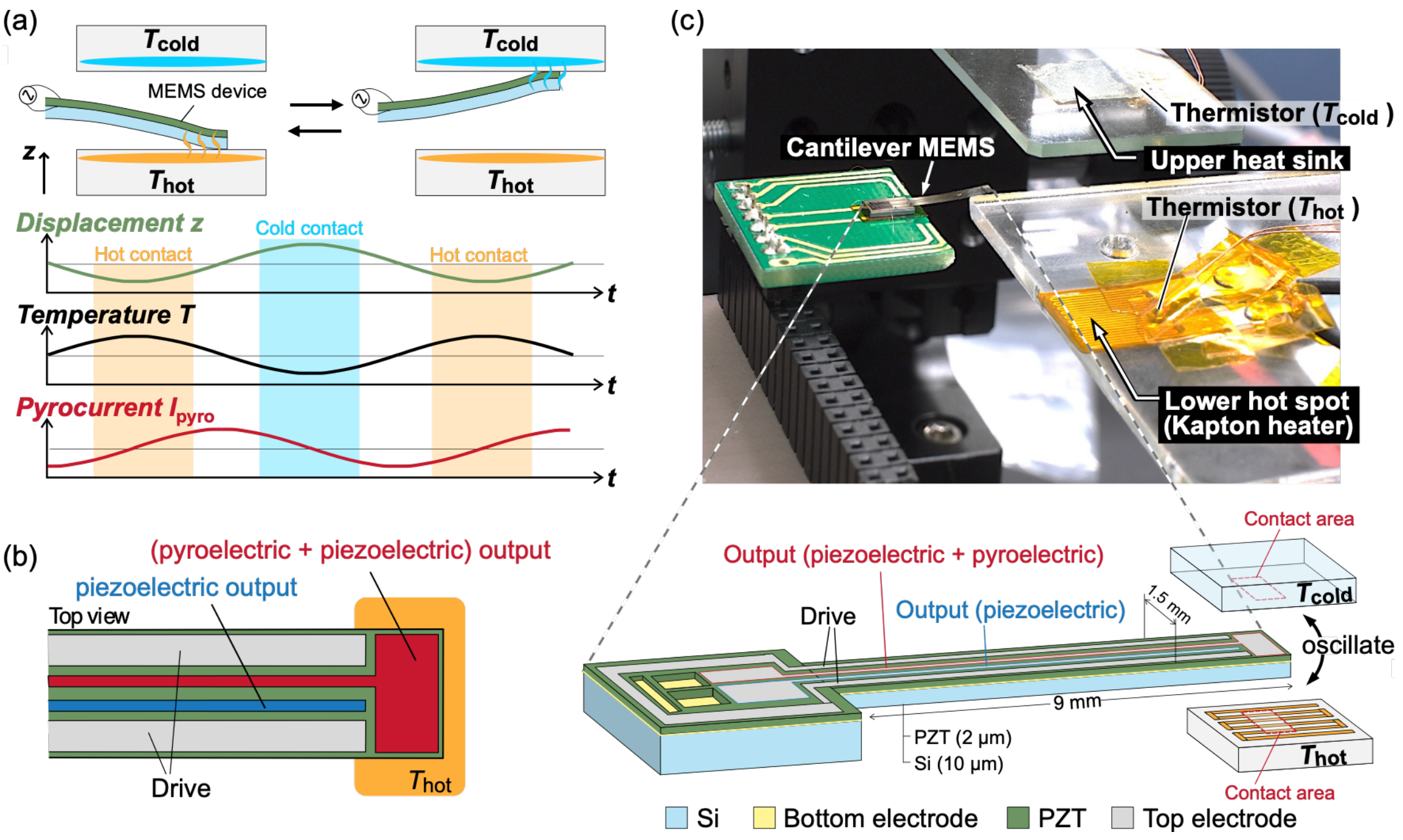

*Figure 2: (a) Operating principle of the active MEMS thermal switch for pyroelectric energy harvesting, showing the sequence from piezoelectric cantilever displacement to periodic temperature modulation and pyrocurrent generation. (b) Electrode configuration for separating the pyroelectric and piezoelectric responses, including the drive electrode, reference piezoelectric output electrode, and (pyroelectric + piezoelectric) output electrode with the thermal contact pad. (c) Photograph and schematic view of the experimental setup, in which the cantilever periodically contacts the lower heat source and upper heat sink.*

Furthermore, by comparing the measured frequency dependence with the response predicted from the independently measured pyroelectric coefficient of the PZT film and combining this comparison with transient thermal simulations, we identify heat transfer across the dynamic contact interface as a key factor limiting high-frequency operation. These results demonstrate the feasibility of extending pyroelectric energy harvesting from steady heat sources toward higher-frequency operation using a dual-function piezoelectric–pyroelectric MEMS architecture, while also revealing the thermal-interface bottleneck that must be addressed for further increases in operating frequency and output power.

## DEVICE CONCEPT AND DESIGN

The proposed device is a ferroelectric/Si cantilever that combines piezoelectric actuation and pyroelectric generation in a single structure (Fig. 2a). Piezoelectric actuation periodically switches the thermal contact between a lower steady heat source and an upper heat sink, generating periodic temperature variations in the ferroelectric layer. These temperature variations simultaneously produce a periodic pyrocurrent, enabling the same ferroelectric layer to function as both the actuator and the energy-conversion element.

Because cantilever bending generates a piezoelectric current that is superimposed on the pyroelectric response, the two output electrodes, in addition to the drive electrode, are designed to separate these contributions (Fig. 2b). One output electrode is connected to a thermal contact pad at the cantilever tip, which periodically contacts the heater and heat sink and therefore experiences a temperature variation. The other output electrode has no thermal contact pad and serves as a reference for the piezoelectric response. The difference is therefore used to extract the pyroelectric contribution,

$$I_{\text{pyro}} \approx I_{(\text{pyro+piezo})} - I_{\text{piezo}}. \quad (1)$$

## EXPERIMENTAL SETUP

The cantilever consists of a 10 μm-thick Si layer, a 2 μm-thick lead zirconate titanate (PZT) layer, and electrodes. The cantilever was fabricated from a commercial PZT capacitor/silicon-on-insulator wafer (I-PEX Inc.). The top electrode, PZT layer, and bottom electrode were sequentially patterned and etched, followed by backside Si etching and removal of the buried oxide layer to release the cantilever.

The cantilever device is placed between a film heater (lower heat source) and a Si chip (upper heat sink), with reservoir temperatures monitored by thermistors (G10K3976B1, TE Connectivity) as shown in Fig. 2c. Figure 3a shows the dependence of the temperature difference between the reservoirs ($\Delta T$) on heater power. $\Delta T$ increases linearly with heater power, confirming that $\Delta T$ can be systematically controlled by the heater power.

The pyroelectric coefficient of the PZT was independently measured to establish the material-level pyroelectric response used for comparison with the device output. Figure 3b shows the DC-bias-dependent pyroelectric coefficient measured directly using periodic heating [9]–[11]. The PZT film exhibited polarization switching with an imprint toward the negative-voltage direction. A large pyroelectric coefficient was already observed in the as-grown state without electrical poling process, consistent with the preferential polarization state of the epitaxial film. The pyroelectric response was well saturated after polarization switching and showed only a slight decrease at higher voltages, which can be associated with the dielectric contribution [12]. Over the 0 to +30 V range used in this study, corresponding to a 15 V DC bias superimposed with 15 V sinusoidal voltage, the average pyroelectric coefficient was approximately $\Pi = -115\ \mu C\ m^{-2}\ K^{-1}$.

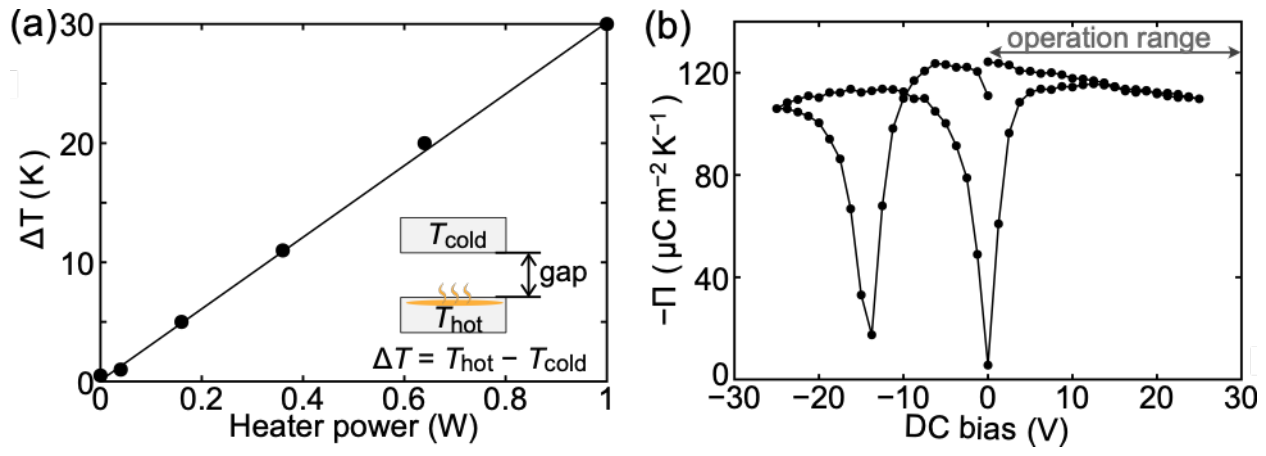

*Figure 3: (a) Measured temperature difference $\Delta T$ between the hot and cold reservoirs as a function of heater power. The inset illustrates the definition of $\Delta T = T_{hot} - T_{cold}$. (b) DC-bias dependence of the directly measured pyroelectric coefficient of the PZT film. The arrow indicates the 0–30 V voltage range used during device operation.*

## RESULTS AND DISCUSSION

Figure 4a shows representative time-domain current signals during operation at 5 Hz. The red and blue curves represent the responses from the (pyroelectric + piezoelectric) output electrode and the piezoelectric output electrode, respectively, with a heater power of 1 W. Their difference therefore provides an estimate of the pyroelectric signal.

This pyroelectric contribution is also supported by the comparison with the heater-off measurement. The black curve shows the signal from the (pyroelectric + piezoelectric) output electrode with the heater turned off. The black curve nearly overlaps with the blue piezoelectric signal, indicating that the additional current observed in the red curve originates mainly from the pyroelectric response induced by periodic thermal contact.

Figure 4b shows the $\Delta T$ dependence of the current amplitudes from the two output electrodes measured during 5 Hz operation. The pyroelectric signal increases with $\Delta T$, as expected. Figure 4c shows the frequency ($f$) dependence of the current amplitudes measured at $\Delta T = 30$ K, demonstrating an enhancement of the pyroelectric signal with increasing frequency over the range of 1–110 Hz, consistent with the fast thermal response of the thin PZT/Si MEMS structure.

We note that the (pyroelectric + piezoelectric) output signal does not completely coincide with the piezoelectric output signal when the heater is turned off. Two possible contributions can account for this difference. First, because the upper heat sink is more exposed to ambient airflow than the lower heat source, a small temperature difference of $\Delta T = 0.5$ K was measured even with the heater off. Second, the thermal contact pad may generate an additional piezoelectric response that is absent from the piezoelectric output electrode.

Next, we quantitatively evaluate the measured current. The pyrocurrent amplitude obtained by subtracting the two output signals is 11.25 nA at 1 Hz and $\Delta T = 30$ K. After subtracting the estimated additional piezoelectric contribution from the thermal contact pad, the pyrocurrent amplitude is 10.25 nA. This value is in reasonable agreement with the calculated amplitude of 12.75 nA obtained from

$$I_{\mathrm{pyro,calc}} = -\Pi A\, dT/dt, \qquad (2)$$

where $A$ is the area of the thermal contact pad, and $dT/dt$ denotes the amplitude of the PZT temperature change rate. Assuming sinusoidal temperature modulation of the PZT between the two reservoir temperatures,

$$dT/dt = 2\pi f \Delta T / 2. \qquad (3)$$

This agreement indicates that the device performs both piezoelectrically driven heat exchange and pyrocurrent generation as designed. Ideally, the pyrocurrent amplitude is expected to increase linearly with frequency according to Eqs. (2) and (3), provided that the PZT undergoes the full temperature excursion between the two reservoirs in each cycle. At higher frequencies, however, the measured current amplitude becomes smaller than the value expected from Eq. (2). At 10 Hz, for example, the measured pyrocurrent amplitude is 36.48 nA, approximately 70% smaller than the calculated value of 127.46 nA.

To clarify the suppression at higher frequencies, the transient thermal response after thermal contact was simulated using COMSOL Multiphysics (Fig. 4d). The temperature evolution in the PZT region after the cantilever contacted the film heater was calculated by varying the area-specific thermal contact resistance between the cantilever and heater, $R_{\mathrm{th,c}}$, from 250 to 4000 mm$^2$ K W$^{-1}$. As the thermal contact resistance increased, the temperature rise became slower, indicating a reduced effective temperature excursion of the PZT during high-frequency operation. In particular, at high operating frequencies, the thermal contact time per cycle becomes shorter than the thermal relaxation time, preventing the PZT from sufficiently approaching the temperatures of the hot and cold reservoirs and thereby reducing its effective temperature modulation.

The experimentally observed suppression of the pyroelectric response at higher frequencies is consistent with the transient thermal response calculated for $R_{\mathrm{th,c}}$ on the order of $10^3$ mm$^2$ K W$^{-1}$ or higher. Although the actual contact conditions, including contact duration, remain uncertain, the half-period at 10 Hz is 0.05 s, and the actual contact time with the hot reservoir should be shorter. Under such conditions, thermal relaxation becomes substantially incomplete for $R_{\mathrm{th,c}} > 1000$ mm$^2$ K W$^{-1}$. These results indicate that high-frequency pyroelectric energy harvesting is limited not only by the mechanical actuation speed and heat conduction within the device, but also by heat transfer across the dynamic contact interface. Therefore, improving the dynamic thermal contact is essential for maintaining a sufficient temperature excursion in each cycle and further increasing the output power at high frequencies.

## CONCLUSIONS

This study demonstrates that active piezoelectric thermal switching enables pyroelectric energy harvesting from a steady heat source over a wide frequency range of 1–110 Hz and at temperature differences down to 5 K. A single PZT film serves as both the piezoelectric actuator and the pyroelectric energy-conversion element, providing a simple integrated structure for generating temperature modulation and pyroelectric energy conversion. The frequency-dependent response and transient thermal simulations further reveal that heat transfer across the dynamic contact interface is a key factor limiting high-frequency operation. These results open a new route for recovering low-grade heat using actively controlled pyroelectric MEMS and provide a design guideline for further increasing the operating frequency and output power.

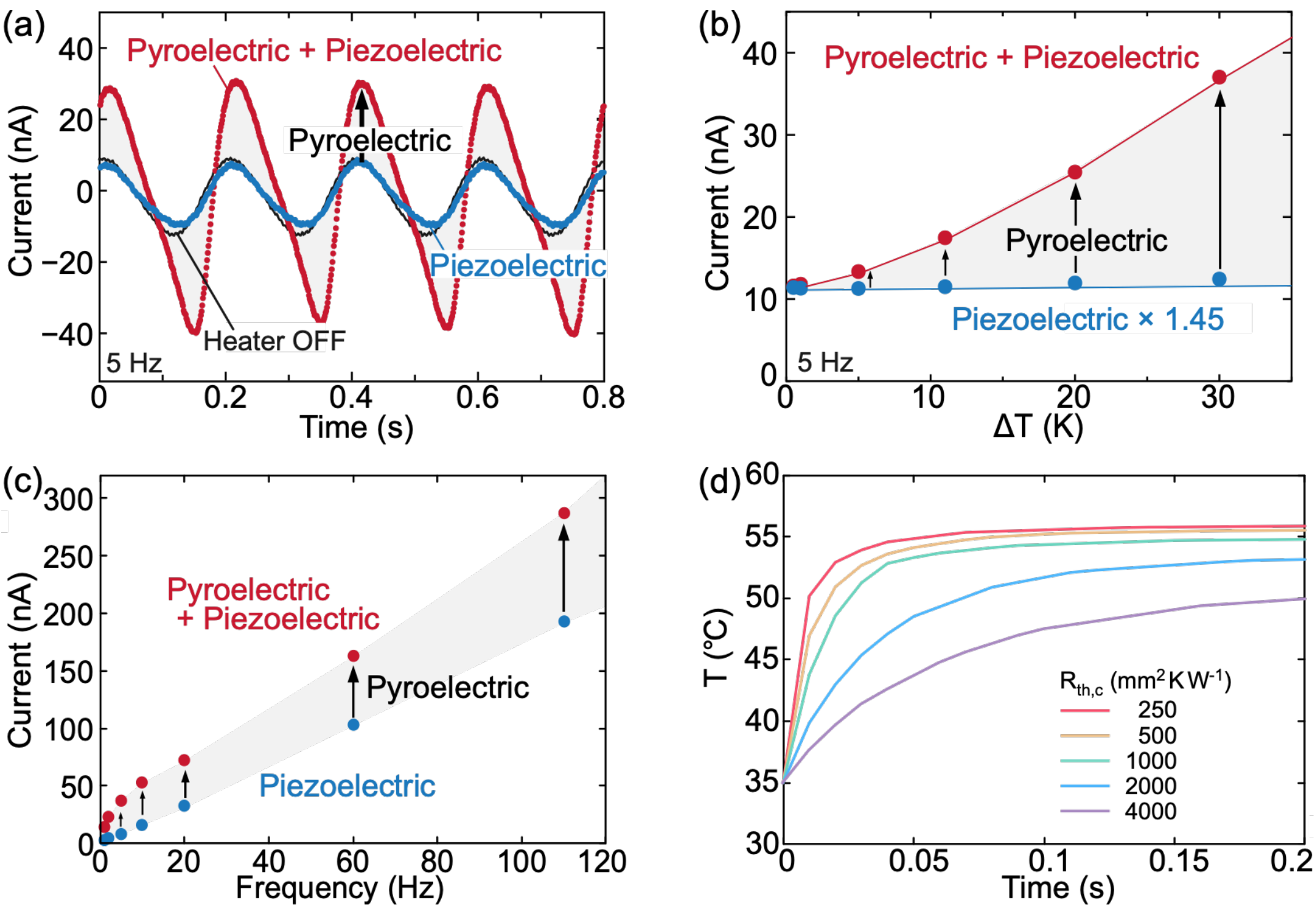

*Figure 4: Device-level pyroelectric response. (a) Time-domain current signals measured at 5 Hz, showing the combined pyroelectric + piezoelectric response (red), reference piezoelectric response (blue), and heater-off response (black). (b) $\Delta T$ dependence of the output amplitudes at 5 Hz. (c) Frequency dependence of the output amplitudes at $\Delta T$ = 30 K. (d) Simulated transient thermal response of the PZT after thermal contact for different thermal contact resistances.*

## ACKNOWLEDGEMENTS

The fabrication in this work was supported by APARCHE at AIST. This work was supported by JST ACT-X (Grant No. JPMJAX23K3) and JSPS KAKENHI (Grant No. JP23K13668, JP25K01644).

## CONTACT

*J.U., tel: +81-8022161934; j-usami@aist.go.jp
*Y.O., tel: +81-8022497903; yuki.okamoto@aist.go.jp